\documentclass[12pt,preprint]{aastex}
\usepackage{apjfonts}
\usepackage{graphicx}

\shorttitle{Simulations of FU Ori} \shortauthors{Zhu et al.}

\begin{document}

\title{2-D simulations of FU Orionis disk outbursts}

\author{Zhaohuan Zhu\altaffilmark{1}, Lee Hartmann\altaffilmark{1}, Charles Gammie \altaffilmark{2,3}, Jonathan C. McKinney \altaffilmark{4}}

\altaffiltext{1}{Dept. of Astronomy, University of Michigan, 500
Church St., Ann Arbor, MI 48105} \altaffiltext{2}{Dept. of
Astronomy, University of Illinois Urbana-Champaign, 1002 W. Green
St., Urbana, IL 61801} \altaffiltext{3}{Dept. of Physics, University
of Illinois Urbana-Champaign} \altaffiltext{4} {Department of
Physics and Kavli Institute for Particle Astrophysics and Cosmology,
Stanford University, Stanford, CA 94305-4060, USA}

\email{zhuzh@umich.edu, lhartm@umich.edu, gammie@illinois.edu,
jmckinne@stanford.edu}

\newcommand\msun{\rm M_{\odot}}
\newcommand\lsun{\rm L_{\odot}}
\newcommand\rsun{\rm R_{\odot}}
\newcommand\msunyr{\rm M_{\odot}\,yr^{-1}}
\newcommand\be{\begin{equation}}
\newcommand\en{\end{equation}}
\newcommand\cm{\rm cm}
\newcommand\gm{\rm g}
\newcommand\kms{\rm{\, km \, s^{-1}}}
\newcommand\K{\rm K}
\newcommand\au{\rm AU}
\newcommand\yr{\rm yr}
\newcommand\etal{{\rm et al}.\ }
\newcommand\sd{\partial}

\begin{abstract}

We have developed time-dependent models of FU Ori accretion
outbursts to explore the physical properties of protostellar disks.
Our two-dimensional, axisymmetric models incorporate full vertical
structure with a new treatment of the radiative boundary condition
for the disk photosphere. We find that FU Ori-type outbursts can be
explained by a slow accumulation of matter due to gravitational
instability. Eventually this triggers the magnetorotational
instability, which leads to rapid accretion. The thermal instability
is triggered in the inner disk but this instability is not necessary
for the outburst. An accurate disk vertical structure, including
convection, is important for understanding the outburst behavior.
Large convective eddies develop during the high state in the inner
disk. The models are in agreement with Spitzer IRS spectra and also
with peak accretion rates and decay timescales of observed
outbursts, though some objects show faster rise timescale. We also
propose that convection may account for the observed
mild-supersonic turbulence and the short-timescale variations of FU
Orionis objects.

\end{abstract}

\keywords{accretion, accretion disks, convection, circumstellar
matter, stars: formation,stars: variables: other, stars: pre-main
sequence} \

\section{Introduction}

The FU Orionis systems are a small but remarkable class of young
stellar objects which undergo outbursts in optical light of 5
magnitudes or more \citep{herbig77}. While the rise times for
outbursts are usually very short ($\sim 1-10 \yr$), the decay
timescales range from decades to centuries. In outbursts, they
exhibit F-G supergiant optical spectra and K-M supergiant
near-infrared spectra dominated by deep CO overtone absorption.  The
FU Ori objects also show distinctive reflection nebulae, large
infrared excesses of radiation, wavelength dependent spectral types,
and ``double-peaked'' absorption line profiles \citep{Lee96}.  The
accretion disk model for FU Ori objects proposed by Hartmann $\&$
Kenyon (1985, 1987a, 1987b) and \cite{kenyon88} can explain the
peculiarities enumerated above in a straightforward manner.

Bell $\&$ Lin (1994) suggested that the thermal instability (TI) was
responsible for FU Ori outbursts. To agree with the outburst mass
accretion rates ($\sim 10^{-4} \msunyr$) and the outburst duration
time ($\sim 100$ yr), with the TI, Bell $\&$ Lin used very low
viscosities both before and after the outbursts. However, this
theory predicts that the outburst is confined to inner disk radii
$\lesssim 0.1 \au$, whereas Zhu et al. (2007, 2008) found that the
high accretion rate region must extend to $\gtrsim 0.5 \au$, based
on modeling {\em Spitzer} IRS data.

\cite{Bonnel1992} suggested that a binary companion on an eccentric
orbit could perturb the disk and cause repeated outbursts of
accretion. However, in the case of FU Ori, no close companion is
known, and no radial velocity variation has been detected
\citep{petrov2008}. Vorobyov $\&$ Basu (2005, 2006) suggested that
gravitational instabilities might produce the non-steady accretion,
but could not carry out the calculation within the inner disk; thus
the details of the accretion event are uncertain. \cite{Lodato2004}
suggested that planets might dam up disk material until a certain
point, resulting in outbursts. This model also depends upon the TI
and assumes a very low viscosity as in Bell $\&$ Lin (1994).

The rediscovery and application of the magnetorotational instability
(MRI) to disk angular momentum transport have revolutionized our
understanding of accretion in ionized disks (e.g. Balbus $\&$ Hawley
1998 and references therein). However, it was recognized that T
Tauri disks are cold and thus likely to have such low ionization
levels that the MRI cannot operate, at least in some regions of the
disk (e.g., Reyes-Ruiz $\&$ Stepinski 1995). Gammie (1996) suggested
that nonthermal ionization by cosmic rays could lead to accretion in
upper disk layers, with a "dead zone" in the midplane. This picture
implies that accretion is not steady throughout the disk, and the
mass should pile up at small radii.

\cite{armitage01} considered the time evolution of layered disks
including angular momentum transport by the gravitational
instability (GI). Within a certain range of mass addition by infall
in the outer disk, they obtained episodic outbursts of accretion at
$\sim 10^{-5}\msunyr$ lasting $\sim 10^{4}$ yr. However, these
outbursts are too weak and last too long compared with FU Ori
outbursts. Gammie (1999) and Book \& Hartmann (2004) were able to
get outbursts more like FU Ori behavior with a different set of
parameters. The Armitage, Gammie, Book \& Hartmann models were
vertically averaged and had schematic treatments of radial energy
transport.

Earlier models of layered disk evolution have been based on
phenomenological one-zone, or two-zone, models for the vertical
structure of the disk.  It is quite difficult to formulate these models
in a consistent way, particularly near transition radii where the
temperature or ionization structure of the disk changes sharply and
radial energy transport may be important.  To make progress it is
necessary to consider more accurate vertical structure and more accurate
energy transport models.

In this paper, we construct two-dimensional (axisymmetric) layered
disk models.  We find that the MRI triggering by the GI can
explain FU Ori-type outbursts if the MRI is thermally activated at
T$_{M}$$\sim$1200 K. In \S 2, the equations and methods of the 2D
models are introduced. In \S 3, outburst behavior is described. In
\S 4, we focus on vertical structure and convection in the
disk. We discuss implications of the 2D simulations compared with
observations in \S 5, and summarize our results in \S 6.

\section{Methods}

In this work we are particularly interested in accurately modeling
the flow of energy within the disk.  A full treatment would
numerically follow the development of MRI, GI, and radiative,
convective, and chemical energy transport in 3D.  As this is not yet
possible, we use the phenomenological $\alpha$-based viscosity to
represent the angular momentum transport by the MRI and GI and
heating by dissipation of turbulence. An accurate treatment of
energy balance (viscous heating and radiative cooling) is essential
to understand the outburst, especially for the thermal instability.
We implemented the realistic radiative cooling in the ZEUS-type code
of \cite{Mckinney2002}.

To test our code, we set a hot region in an optically thick disk
region and then let the radiative energy diffuse without evolving
the hydrodynamics. The hot region expands nicely in a round shape in
R-Z space with the correct flux. We have also tested the total
energy conservation during our outburst simulations, finding that
the energy deficit during outburst is only $<$~1\% of the total
energy radiated.

\subsection{Viscous heating}

The momentum equation and energy equations are
\begin{equation}
\rho\frac{D v}{Dt}=-\nabla p-\rho\nabla \Psi-\nabla\cdot \Pi \,,
\end{equation}
\begin{equation}
\rho\frac{D}{Dt}\left(\frac{e}{\rho}\right)=-p \nabla\cdot v
+\Phi-\Lambda \,.
\end{equation}
Here, as usual, $D/Dt\equiv \partial/\partial t+v \cdot \nabla$ is
the Lagrangian time derivative, $\rho$ is the mass density , $v$ is
the velocity, $e$ is the internal energy density of the gas, $p$ is
the gas pressure, $\Psi$ is the gravitational potential, and
$\Lambda$ is the radiative cooling rate. The dissipation function
$\Phi$ is given by the product of the anomalous stress tensor $\Pi$
with the rate-of-strain tensor $\xi$,
\begin{equation}
\Phi=\Pi_{ij}\xi^{ij}
\end{equation}
(sum over indices), where the anomalous stress tensor is the
term-by-term product
\begin{equation}\label{Pi}
\Pi_{ij}=2\rho\nu_{V} \xi_{ij}S_{ij} \label{eq:stress}
\end{equation}
(no sum over indices), where $S_{ij}$ is a symmetric matrix filled
with 0 or 1 that serves as a switch for each component of the
anomalous stress. In current simulations, we switched on all the
component of the anomalous stress. It is not yet known whether any
prescription for $\Pi_{ij}$ is a good approximation for modeling MHD
turbulence. We set the viscosity coefficient to
$\nu_{V}=\alpha(c_{s}^{2}/\Omega)sin^{3/2}\theta$, which vanishes
rapidly near the poles \citep{Mckinney2002}. $\alpha$ is the
viscosity parameter and $\Omega$ is the angular velocity at radius
$R$.

The equation of state is
\begin{equation}
p=\left(\gamma-1\right)e\,.
\end{equation}
and the sound speed is
\begin{equation}
c_{s}^{2}=\gamma\frac{k_{B}T}{\mu}
\end{equation}
where $\gamma$ and $\mu$ are the adiabatic index and the mean
molecular weight.  We assume $\gamma=5/3$ and $\mu=2.4 M_{P}$ where
$M_{P}$ is the mass of one proton. The adiabatic index and mean
molecular weight will change at different temperatures and pressure
(especially during outburst when the disk is fully ionized) and a
fully self-consistent treatment would use a tabulated equation of
state. We will discuss a more realistic equation of state in a later
paper.

\subsection{Radiative cooling}

In an accretion disk, radiation is the main cooling mechanism, and
radiation energy and pressure can affect the disk dynamics. Numerically,
there are two different approaches to treat the radiative transport in
an accretion disk.  One is to solve the full radiative transfer equation
\begin{equation}
\rho\frac{D}{Dt}\left(\frac{E}{\rho}\right)=-\nabla\cdot F-\nabla
v:P +4\pi\kappa_{P}B-c\kappa_{E}E\label{eq:ree}
\end{equation}
with the material energy equation:
\begin{equation}
\rho\frac{D}{Dt}\left(\frac{e}{\rho}\right)=-p\nabla\cdot
v-4\pi\kappa_{P}B+c\kappa_{E}E \label{eq:mee}
\end{equation}
in both the optically thick and thin regions by an implicit method
\citep{turner01,wunsch,Hirose2006}.  Here, as above, $\rho$,$e$,$v$,
and $p$ are the gas mass density, energy density, velocity and
scalar isotropic pressure respectively, while E,F, and P are the
total radiation energy density, momentum density and pressure
tensor, respectively. $\kappa_{P}$ and $\kappa_{E}$ are Planck mean
and energy mean absorption coefficients. An implicit method is
required because of the extremely short time step needed by the
explicit method (grid size over the speed of light), but the
implicit method is time consuming, requiring solution of an
N$^{2}\times$N$^{2}$ matrix equation at each timestep.

A second, less accurate but less costly, approach is to add a
radiative source (heating/cooling) term in the energy equation, use
the diffusion approximation in the optically thick region, and use
an approximation to treat the optically thin region
\citep{Boss2002,Boley2006,Cai2008}.  When the radiation pressure is
negligible, as in the case for protostellar disks, and the radiation
field is nearly in equilibrium ($D(E/\rho)/Dt=0$), equations
\ref{eq:ree} and \ref{eq:mee} become
\begin{equation}
\rho\frac{D}{Dt}\left(\frac{e}{\rho}\right)=-p\nabla\cdot v-\Lambda
\label{eq:transfer}
\end{equation}
where $\Lambda=\nabla\cdot F$ and $F$ is the radiative flux.

We examine the justification for ignoring the advection term (within
the term on the left side of Eq. \ref{eq:ree}) and the term
describing work done by radiation on the gas (second term of the
right side of Eq. \ref{eq:ree}) following the criterion given by
\cite{1984frh..book.....M}. If in one system, $L$ is the
characteristic size, u is the characteristic velocity and $\tau$ is
the optical depth of $L$,  the terms describing the advection and
the work done by the radiation can be ignored if $\tau\gg$1 and
$\tau\times u/c\ll$1, which is called static diffusion limit. This
criterion can be simply understood as follows:

The timescale for radiation to propagate across a region of length
$L$ and mean optical depth $\tau$ is of order
\begin{equation}
t_E \sim L \tau / c \,.
\end{equation}
For comparison, the material crossing time of this region is
\begin{equation}
t_u \sim L / u\,,
\end{equation}
so that the ratio of these two timescales is
\begin{equation}
{t_E \over t_u} \sim {u \over c} \tau\,.
\end{equation}
If this ratio is far less than one, the region is in radiative
quasi-equilibrium before the material carrying radiation energy away
so that the energy emitted and absorbed equals to the diffusion term
(Eq. \ref{eq:ree}). In a 3-D accretion disk, this typical velocity
is the Keplerian velocity. However in our 2-D axisymmetric
simulation, this velocity is just the mean radial velocity of the gas. Thus,
$t_{u}$ is close to the viscous time which is longer than the
time for sound to cross a scale height $H$ by a factor of order
$\alpha^{-1} (R/H)^{2}$.

In our simulations, even at the maximum temperatures achieved in the
innermost disk, $u/c \lesssim 10^{-5}$ (\S 4) and so the
approximation of radiative equilibrium should be valid except at
very large optical depths.  It turns out that, during the outburst,
vertical Rosseland mean optical depths $\sim 10^4$ in the innermost
disk. Thus our approximation is justified, though it may break down
at smaller radii than we have considered here.

In detail, the radiative transfer in the optically thick disk,
defined as the region where $\tau>\tau_{thick}$, is given in the
diffusion approximation by
\begin{equation}
\Lambda=\triangledown\cdot F,
\end{equation}
where the flux $F$ is
\begin{equation}
F=-\frac{16\sigma\beta T^{3}}{\chi_{R}\rho}\triangledown T,
\label{eq:diffusion}
\end{equation}
and $\beta$ is the flux limiter,
\begin{equation}\label{beta}
\beta=\frac{2+y}{6+3 y+y^{2}}\,,
\end{equation}
with
\begin{equation}
y=\frac{4}{\chi_{R}\rho T}\mid \triangledown T\mid\,.
\end{equation}
where $\chi_{R}$ is the Rosseland mean of the absorption
coefficient.

The optically thin region is more complicated.
In this region we restrict our considerations to radiative transport in
the vertical direction only.  We have developed
two methods to solve this problem in collaboration with members of the
Indiana disk group (D. Durisen, K. Cai, \& A. Boley, 2008, personal
communication).
The first method considers the cooling rate of every cell in the
optically thin region can be written as:
\begin{equation}\label{2}
\Lambda=\rho\kappa_{P}(T)\left(4\sigma T^{4}-4\pi
J(Z)\right)-\rho\kappa_{P}(T_{e})\sigma T_{e}^{4}e^{-\tau_{e}(Z)}\,
\end{equation}
where $\kappa_{P}(T)$ is the Planck mean opacity at the temperature
$T$, $T_{e}$ is the temperature of the envelope, and also J(Z) and
$\tau_{e}(Z)$ are the mean intensity and the optical depth of the
envelop at height Z in the disk atmosphere. The first term is the
energy emitted by the cell, while the second and third terms are the
energy absorbed from the optically thick region and from the
envelope. Using the two stream approximation,
\begin{equation}
J(Z)=\frac{1}{4\pi}\int I d\Omega=\frac{1}{4\pi}\left(2\pi
I_{+}+2\pi I_{-}\right).
\end{equation}
This treatment assumes flux conservation and
thus ignores energy dissipation in the atmosphere.
The net flux $F_{b}$ from the disk photosphere is then
\begin{equation}
F_{b}=\int I \, \mu \, d\Omega=\pi\left(I_{+}-I_{-}\right)\,,
\end{equation}
where $\mu$ is the cosine of the angle measured from the vertical.
Thus
\begin{equation}
J(Z)=\frac{1}{2}\left(I_{+}+I_{-}\right)=\frac{F_{b}}{2\pi}+I_{-}(Z).
\end{equation}
$F_{b}$ is approximately
\begin{equation}\label{3}
F_{b}=\frac{4\sigma
\left(T_{b}^{4}-T_{e}^{4}\right)}{3\left(\tau_{b}+2/3\right)}
\end{equation}
where $T_{e}$ is a term which accounts for the external irradiation
(see equation (10) in \cite{Boley2006}). $T_{b}$ and $\tau_{b}$ are
the temperature and optical depth of the first optically thick cell
 right below the photosphere (again, see \cite{Boley2006}).Finally, we must
set $I_{-}(Z)$ to obtain a closed expression for $\Lambda$:
\begin{equation}
I_{-}(Z)= \int_{Z}^{\infty}\rho\kappa_{P}\frac{\sigma
T(z)^{4}}{\pi}dz+\frac{\sigma T^{4}_{env}e^{-\tau_{en}(Z)}}{\pi}\,.
\end{equation}
which is the intensity from the optically thin region above height
$Z$. Theoretically, the first term should also have e$^{-\tau}$.
However, since we only consider the optically thin region here, we
ignore it to simplify the calculation and save computational time.

The second method is much simpler but limited to this particular
problem with no irradiation. As we do not consider irradiation, we
can use the usual gray atmosphere result. Without the third term in
equation (\ref{2}), we have
\begin{equation}
\Lambda=\rho\kappa_{P}(T)\left(4\sigma T^{4}-4\pi J(Z)\right)\,,
\end{equation}
and in a gray case
\begin{equation}
J=\frac{\sigma}{\pi}
T^{4}(Z)=\frac{\sigma}{\pi}\left(\frac{3}{4}T_{eff}^{4}(\tau+\frac{2}{3})
\right)\,, \label{eq:diffusionT}
\end{equation}
where $T_{eff}$ is given by the boundary flux $F_{b}=\sigma
T_{eff}^{4}$ as equation \ref{3}. In detail, counting $\theta$ grid
cell index i as increasing with increasing $\theta$ and $\tau$, we
set the first grid with $\tau (R, \theta_i) > 3$ as the boundary
grid and $\tau_{b}$=$\tau (R, \theta_i)$. Then we derive the
effective temperature (or equivalently, the boundary flux) using the
gray approximation
\begin{equation}
F_b=\sigma T_{eff}^4 = \frac{4\sigma T_{b}^{4}}{3(\tau_{b}+2/3)} \,.
\end{equation}
With $T_{eff}$, we use equation \ref{eq:diffusionT} to set the
temperatures of all cells at $R$, $\theta < \theta_i$, and use the
diffusion equation (\ref{eq:diffusion}) at all $r, \theta \geq
\theta_{i}$.  This way of matching the "optically thin" (gray)
atmosphere to the "optically thick" (diffusion equation) interior
removes the temperature jump seen at the boundary between the two
treatments seen in the simulations of \cite{Cai2008}.

Both methods give similar results if there is no irradiation(see
Appendix A) - basically, they are the same except for the assumed
directionality of the radiation field. We use the second method in
this paper, although the first method is of interest for cases in
which external irradiation is important.

We have adopted the Bell \& Lin (1994) opacity fits to facilitate
comparison with previous calculations \citep{bell94,armitage01};
subsequent improvements in low-temperature molecular opacities
will modify results at low temperatures (Zhu et al 2009), but should
not substantially affect the outburst behavior driven at higher temperatures.

\subsection{Angular Momentum Transport}

To complete the model we need a prescription for the dimensionless
viscosity $\alpha$.  The key physics we want to capture is the
dependence of the MRI and GI transport on position in the disk
(see, e.g., Miller \& Stone 2000 for MRI activity as a function of height
in the disk).

 For the MRI, we set $\alpha_M = 0.1$ when $T > T_M$, where $T_M$
is a critical temperature (typically $1200\K$) where thermal
ionization makes the disk conductive enough to sustain the MRI.
\footnote{3D MHD simulations suggest that the energy dissipation
rate per unit mass is slightly higher at $\sim 2 H$, which means
$\alpha$ is not constant with height} In addition we calculate
$\Sigma_+ = \int_0^\theta R d\theta' \rho$ (and in models where the
disk is not assumed to be symmetric about the midplane, $\Sigma_- =
\int_\theta^\pi R d\theta'\rho$) and set $\alpha_M = 0.1$ if
$\Sigma_+$ or $\Sigma_- < \Sigma_{crit}$, where $\Sigma_{crit}$ is
the maximum surface density of the MRI active layer on one side of
the disk (In our current model we assume $\Sigma_{crit}$=50 g
cm$^{-2}$ ). Otherwise the disk is magnetically 'dead' and $\alpha_M
= 0$. Also to avoid the strong MRI viscous heating at the disk's
upper photosphere with very low density, we set $\alpha=0$ for the
region with $\tau<10^{-3}$.

The value of $\alpha_{M}$ predicted by theory is uncertain, with
widely-varying estimates depending upon mean magnetic fluxes,
magnetic Prandtl numbers, and other assumptions (Guan et al. 2008,
Johansen \& Levin 2008, Fromang et al. 2007).  Given these
uncertainties, observational constraints take on special
significance. A recent review of the observational evidence by King,
Pringle, \& Livio (2007) argues that $\alpha_M$ must be large, of
the order 0.1-0.4, based in part on observations of dwarf novae and
X-ray binaries where there is no question of gravitational
instability. Our empirical study of FU Ori, where we were able to
constrain the radial extent of the high accretion state, implies
values of $\alpha \sim 0.1$. (Zhu \etal 2007).

For the GI, we calculate the Toomre $Q$ parameter
\begin{equation}
Q = {c_s \Omega \over \pi G \Sigma} \,.
\end{equation}
When $Q \sim 1$ the disk is gravitationally unstable.  Full global
models \citep{Boley2006} and physical arguments (Gammie 2001) suggest
that if the disk is sufficiently thin angular momentum transport by GI
is localized and can be crudely described by an equivalent $\alpha$.  We
set
\begin{equation}
\alpha_{Q}(R,\theta)=e^{-Q(R)^{2}} \label{eq:alphaQ},
\end{equation}
which is independent of $\theta$.  This has the sensible effect of
turning on GI driven angular momentum transport when $Q < 2$ and
turning off GI driven angular momentum transport when $Q$ is large.
In regions with both GI and MRI we set $\alpha = \alpha_M$ (in these
regions $\alpha_M \gg \alpha_Q$).

The maximum $\alpha_{Q}$ is $<$0.1 during the simulations,
consistent with our assumption that fragmentation does not occur.
Our prescription for $\alpha_{Q}$ on Q is heuristic and other forms
are plausible.  Gammie (2001) and Lodato \& Rice (2004, 2005) found
that the amplitude of GI depends on the cooling rate in addition to
Q; and Balbus \& Papaloizou (1999) argued that the angular
momentum flux carried by GI is a global effect and the local
viscosity description can only be valid near the corotation radius.
However, using self-consistent cooling, Boley et al. 2006 showed
that the $\alpha_{GI}$ agrees with the Gammie description very well,
which assumes that the GI viscous heating is locally balanced by the
cooling. Recent smooth particle hydrodynamics SPH simulations by
Cossins et al. (2009) also showed that GI disks with tightly wound
spiral arms can be approximated by a local treatment.

\section{2D results}

We use spherical polar coordinates and assume axisymmetry in all our
simulations. To sample both the inner disk and the outer disk, we
use a logarithmic radial grid (radial grids are uniform in
log(R-R$_{in}$)) from 0.06-40 AU and a uniform $\theta$ grid in
$(0.5 \pm 0.15) \pi$. This gives adequate radial resolution for both
the inner and outer disk, and spans a large enough range in $\theta$
to include the optically thick region of the disk.  Parameters for
all the 2D simulations are listed in Table 1.

\subsection{Initial and Boundary Conditions}

All our disk models begin with the initial temperature distribution
\begin{equation}
T(R,\theta) = 300 (R/AU)^{-1/2} \K,
\end{equation}
and $\rho = \Sigma(R)/(H {2 \pi}^{1/2}) \exp(-R^2\theta^2/(2 H^2))$,
$H \equiv c_s/\Omega$ from 0.4-100 AU \footnote{At this stage we use
larger radii and lower gird resolution than Table 1 to make the disk
quickly evolve to the state just before the outburst is triggered.
When the outburst is about to be triggered, we interpolate the grid
to get a higher resolution grid from 0.06-40 AU.}, so the disk is
approximately in dynamical equilibrium. We set $\Sigma(R)$ using a
$\dot{M}$=10$^{-5}$ M$_{\odot}$/yr constant $\alpha$ ($\alpha=0.1$)
one-zone accretion disk model (see Equation 3 in \cite{whitney}).
Although this is not an exact steady-state model, the disk relaxes
rapidly at the beginning of the simulation.

We apply outflow boundary conditions \citep{Stone1992} to all
boundaries.  In some simulations the disk is assumed to be symmetric
about the midplane to reduce computational cost; for these half-disk
simulations we use a reflecting boundary condition \citep{Stone1992}
at the midplane. The inner boundary is set to either 0.06 AU, 0.1
AU, or 0.3 AU to examine how different inner radii affect disk
evolution. These parameters are listed in Table 1.

We follow the evolution of the disk for 10$^{5}$ years.  Typically
several outbursts occur during the simulation.  After the first
outburst, the subsequent outbursts are similar and do not depend on
the initial condition. Thus, we use the disk density and temperature
distribution after the first two outbursts as the new initial
condition to study the third outburst for various 1D and 2D models.
The disk mass for this initial condition is $\sim$ 0.5 $\msun$.

\subsection{Outbursts}

In the protostellar phase, the disk is unlikely to transport mass
steadily from $\sim$100 AU all the way to the star at an accretion rate
matching the mass infall rate $10^{-6}-10^{-4}\msunyr$ from the envelope
to the outer disk. The outburst sequence we find here is qualitatively
similar to that found by \cite{armitage01}, Gammie (1999), and Book \&
Hartmann (2001). Mass added to the outer disk moves inwards due to GI,
but piles up in the inner disk as GI becomes less effective at small R.
Eventually, the large $\Sigma$ and energy dissipation leads to enough
thermal ionization to trigger the MRI. Then the inner disk accretes at a
much higher mass accretion rate, which resembles FU Orionis-type
outbursts.  After the inner disk has drained out and becomes too cold to
sustain the MRI, the disk returns to the low state.  During the low
state mass continuously accretes from large radius (or from an infalling
envelope) causing matter to pile up at the radius where GI becomes
ineffective yet again, leading to another outburst.

\begin{figure}
\epsscale{.80} \plotone{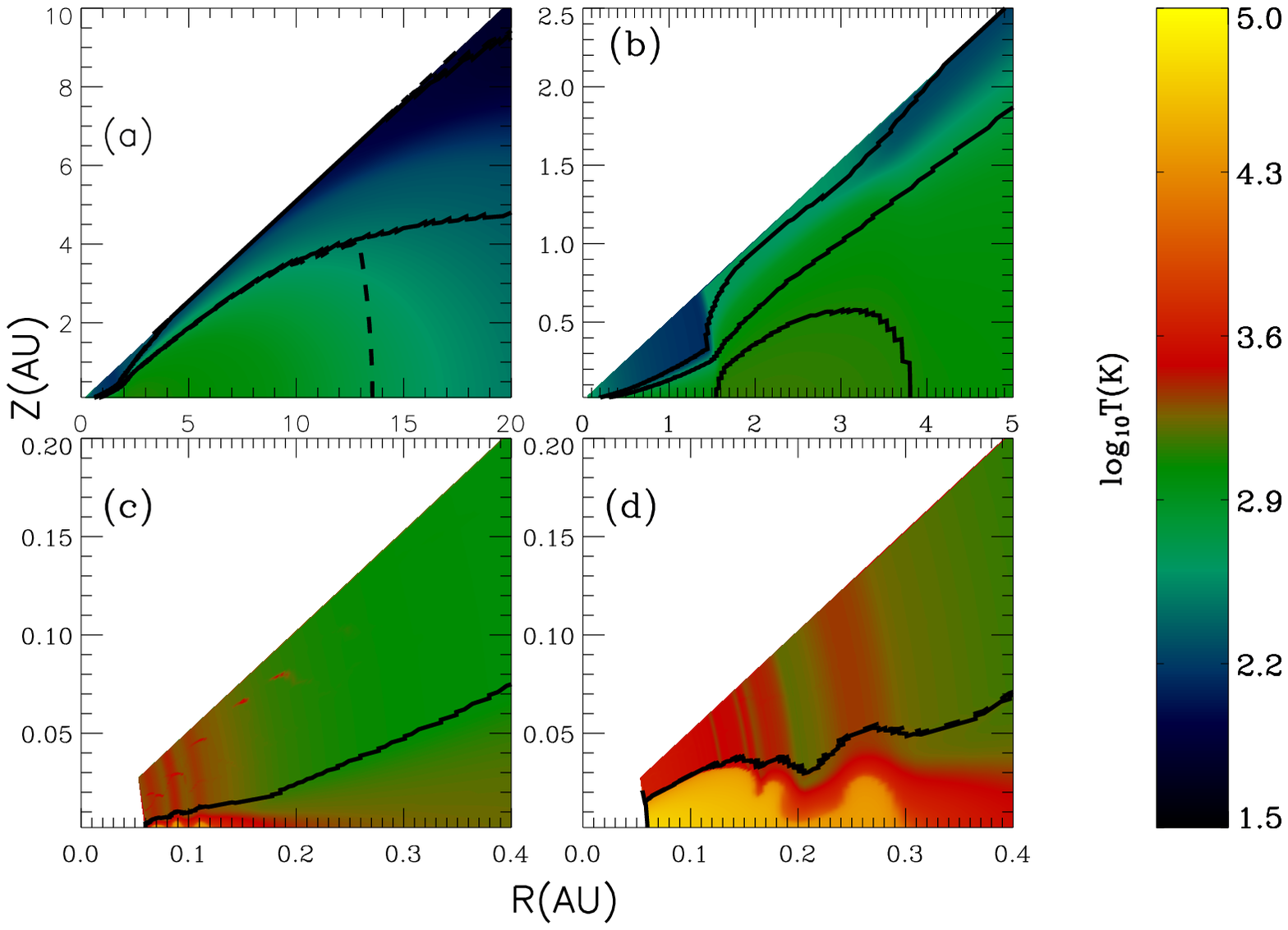}  \caption{ The disk's temperature
color contours at four stages during the outburst in R-Z coordinate.
The black curves are the $\alpha$ contours which also outlines
different regions in the disk. The vertical dashed line in Panel (a)
shows $\alpha_{Q}=0.025$. (a) is the stage before the outburst. (b)
represents the stage when the MRI is triggered at 2 AU. (c)
represents the moment when the TI is triggered at the innermost
region. (d) is the stage when the TI is fully active within 0.3
AU.}\label{fig:5polar}
\end{figure}

\begin{figure}
\epsscale{.80} \plotone{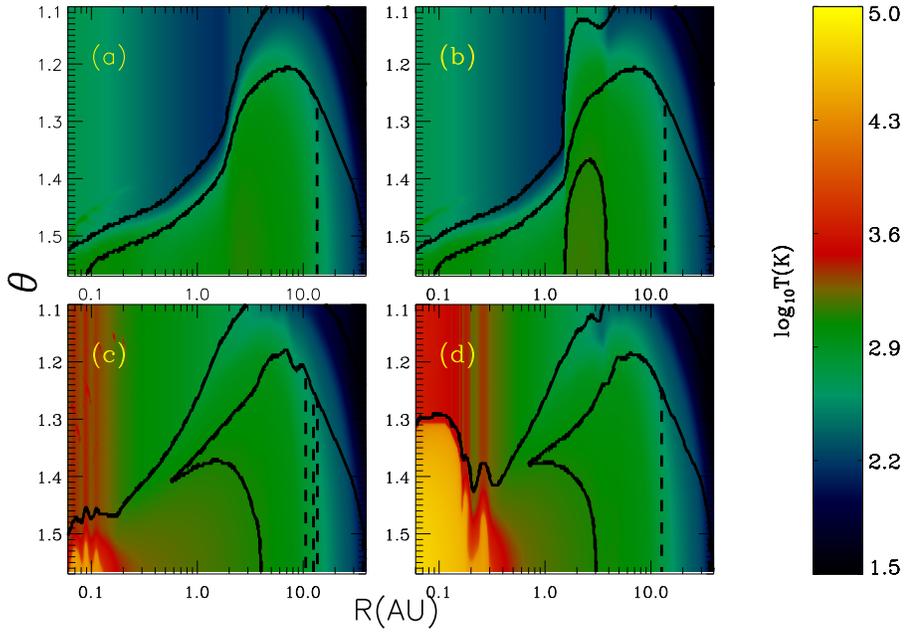}  \caption{ The disk's temperature
color contours at four stages during the outburst, the same as Fig.
\ref{fig:5polar}. The black curves are the $\alpha$ contours which
also outlines different regions in the disk: the uppermost
atmosphere, active layer, and the dead zone close to the midplane.
The dashed line shows $\alpha_{Q}=0.025$. (a) is the stage before
the outburst. (b) represents the stage when the MRI is triggered at
2 AU. (c) represents the moment when the TI is triggered at the
innermost region. (d) is the stage when the TI is fully active
within 0.3 AU.}\label{fig:5}
\end{figure}

\begin{figure}
\epsscale{.80} \plotone{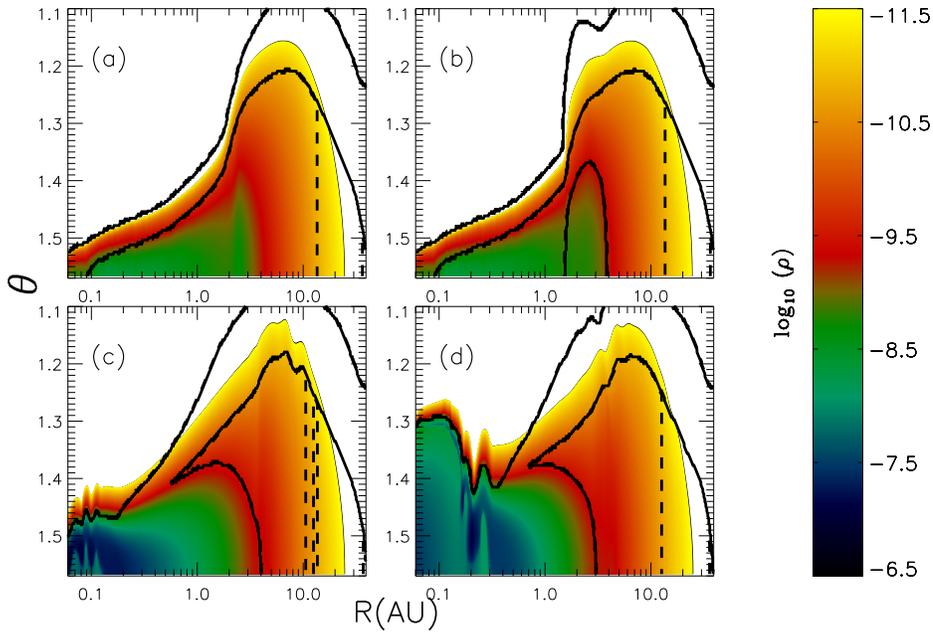} \caption{ Similar to Figure
\ref{fig:5} but shows the density contours. The high density white
regions close to the midplane in (c) and (d) correspond to the
region at the TI low state in Figure \ref{fig:5}. Because it is much
colder than the midplane which is at the TI high state, it needs a
high density to balance the pressure.}\label{fig:5rho}
\end{figure}

\begin{figure}
\epsscale{.80} \plotone{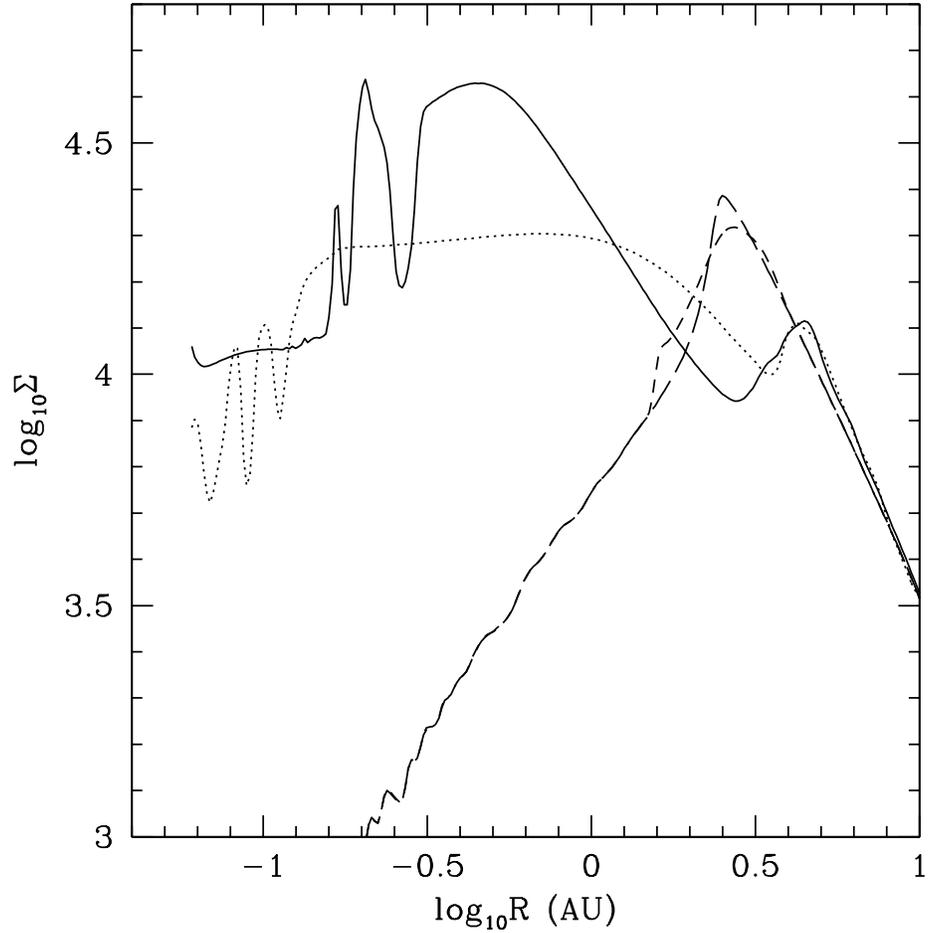} \caption{ The distribution of the
vertically integrated surface density at four corresponding stages
as shown Figure \ref{fig:5}. The long dashed line is the stage
before the outburst, while the short dashed line represents the
stage when the MRI is triggered at 2 AU. The dotted line shows the
moment when the TI is triggered. Finally, the solid line shows the
TI fully active stage.
 }\label{fig:sigma}
\end{figure}

\begin{figure}
\epsscale{.80} \plotone{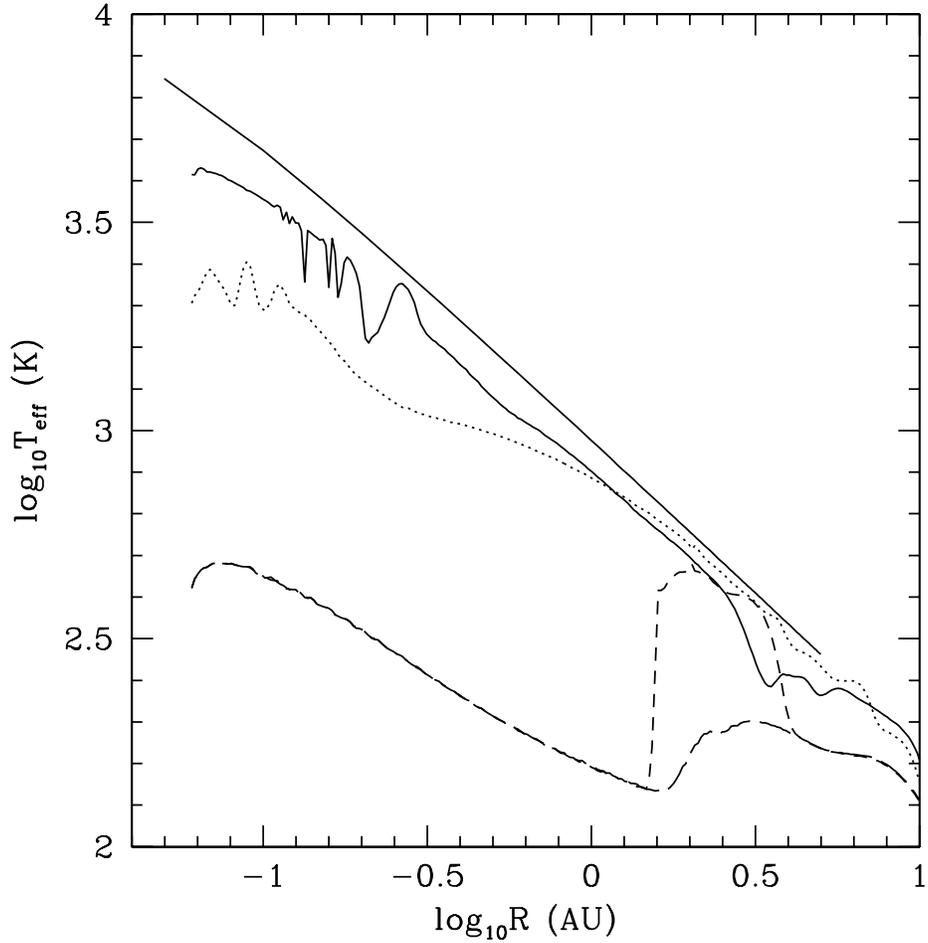} \caption{ Similar to Figure
\ref{fig:sigma} but shows the effective temperature distribution at
the four stages. The long dashed line is the stage before the
outburst, while the short dashed line represents the stage when the
MRI is triggered at 2 AU. The dotted line shows the moment when the
TI is triggered. The solid line shows the TI fully active stage,
while the uppermost solid line shows the effective temperature of a
steady accretion disk with $\dot{M}\sim$2$\times$10$^{-4}\msunyr$
and R$_{in}\sim$5$R_{\odot}$ }\label{fig:teff}
\end{figure}

Figures \ref{fig:5polar}- \ref{fig:teff} show the disk temperature
and density structure at four consecutive stages of the outburst for
our fiducial model. Figure \ref{fig:5polar} is shown in R-Z
coordinate while figures \ref{fig:5} and \ref{fig:5rho} are shown in
log$_{10}$R-$\theta$ coordinates (R from 0.06 to 40 AU, $\theta$
from 1.1 to $\pi/2$). Since the GI, MRI and TI are taking place at
quite different radii, most of our plots are in log$_{10}$R-$\theta$
coordinates so that all of them can be seen in one figure, although
it is less intuitive than R-Z coordinates. The black solid curves
plotted on top of the color contours delimit regions of differing
$\alpha$. Regions above the top solid curve limit the low-$\rho$
photosphere with $\tau<10^{-3}$ to $\alpha=0$. Between the top and
next lower solid curve lies the MRI-active layer with
$\Sigma_{+}$<50 \gm \cm$^{-1}$ and $\alpha=\alpha_{M}=0.1$. Regions
below the two solid curves constitute the "dead zone" with only
$\alpha_{Q}$ if any. The vertical dashed lines at large radii
represent $\alpha_{Q}=0.025$.

Panel (a) of Figures 1-3 show the disk temperature structure before
the outburst. The disk is too cool to sustain any MRI except in the
active layer. The contours in Figures
\ref{fig:5polar}-\ref{fig:5rho} show the outer disk (R>1AU) is
geometrically thick, with a ratio of vertical scale height H to
radius H/R$\sim$0.2, and massive, while the inner disk is thin with
a low surface density.  The inner disk is stable against GI and
cannot transport mass effectively. The mass coming from the GI
driven outer disk piles up at several AU.  The GI thus gradually
becomes stronger at several AU and the disk heats up.

The disk midplane temperature at 2 AU eventually reaches the MRI
trigger temperature (T$_{M}\sim$1200 K) and MRI driven angular
momentum transport turns on.  Panel (b) of Figures
\ref{fig:5polar}-\ref{fig:5rho} show the stage when the thermal-MRI
is triggered at the midplane around 2AU. The thermal-MRI active
region are outlined by the semicircular $\alpha$ contour around 2 AU
in these figures. After the MRI is activated, $\alpha$ increases
from $\alpha_{Q}(2 AU)\sim 0.01$ to $\alpha_{M}=0.1$, which leads to
a higher mass accretion rate. Afterwards, the MRI active region
expands both vertically and radially until the whole inner disk
becomes fully MRI active.

As more and more mass is transferred to the inner disk, the
temperature continues to rise.  At $T \sim 4000 \K$ the disk becomes
thermally unstable and $T$ rises rapidly to $\sim 2 \times 10 ^{4}
\K$. Panel (c) of Figures \ref{fig:5polar}-\ref{fig:5rho} show the
moment the TI is initiated in the innermost disk.  With the
temperature leap, the pressure of the TI active region also
increases significantly. Thus the TI active region expands and
compresses the TI non-active region, leading to a lower density of
the TI active region than nonactive region (lower density at
midplane around 0.1 AU in panel (c) of Fig. \ref{fig:5rho}).

Panel (d) of Figures \ref{fig:5polar}-\ref{fig:5rho} show that after
the TI is triggered in the innermost disk, the TI front travels
outwards to $\sim$ 0.3 AU until the entire inner disk is in a ''high
state.'' The hot inner disk (R<0.2 AU) is puffed up to
H/R$\sim$0.15. The mass accretion rate is $2\times 10^{-4}\msunyr$
at this stage. The central temperature has a sharp drop at 0.3 AU,
beyond which the disk is in a thermally stable state at $T \simeq
3000 \K$. Details of the TI are discussed in Appendix B. At this
transition from the high to low state around $0.2 \au$, some
temperature and density fluctuations are present, which results from
convection. We will discuss convection in detail in \S 5.

The TI high state (T$_{c}$>10,000 K, panel (d) of Fig.
\ref{fig:5polar}-\ref{fig:5rho}) lasts for $\sim$10$^{2}$ years,
draining the inner disk (R < 0.5 AU) until it is not massive and hot
enough to sustain the TI high state; it then returns to the TI low
state. The inner disk may temporarily return to the high state due
to MRI-driven mass inflow but eventually mass is drained from the
MRI active region, it becomes MRI inactive, and the disk returns to
the state shown in panel (a).

Figure \ref{fig:sigma} shows the vertically integrated surface
density at the above stages. Before the outburst, the surface
density peaks around 2 AU (long-dash curve); this radius divides the
inner, GI stable disk from the outer GI unstable disk.  Once the MRI
is triggered at 2 AU, it transports mass into the inner disk
effectively. At the moment the TI is triggered (dotted curve), the
surface density of the inner disk is already one order of magnitude
higher than that before the MRI is triggered. During the TI fully
active stage (solid curve), the relatively low surface density
region within 0.3 AU corresponds to the TI high stage with
T$_{c}$>10$^{4}$ K. The density fluctuations around 0.3 AU result
from convection at the transition between the TI high and low states
as seen in Figures \ref{fig:5polar} and \ref{fig:5rho}. We discuss
the details of convection in \S 4.3

Figure \ref{fig:teff} shows the effective temperature distributions
at the same stages as above. When the MRI is triggered, the
effective temperature increases dramatically with the increasing
accretion rate. Eventually the whole inner disk almost accretes at a
constant rate and the effective temperature is similar to that of a
steady accretion solution at $M\dot{M}=1.8\times10^{-4}M_{\odot}^{2}
yr^{-1}$ (smooth solid curve).  The discrepancy at small radii may
be caused by the outflow inner boundary condition of the 2D
simulation, which means all the velocity gradients relative to the
coordinates (R, $\theta$) are 0.

\subsection{Resolution Study}

\begin{figure}
\epsscale{.80} \plotone{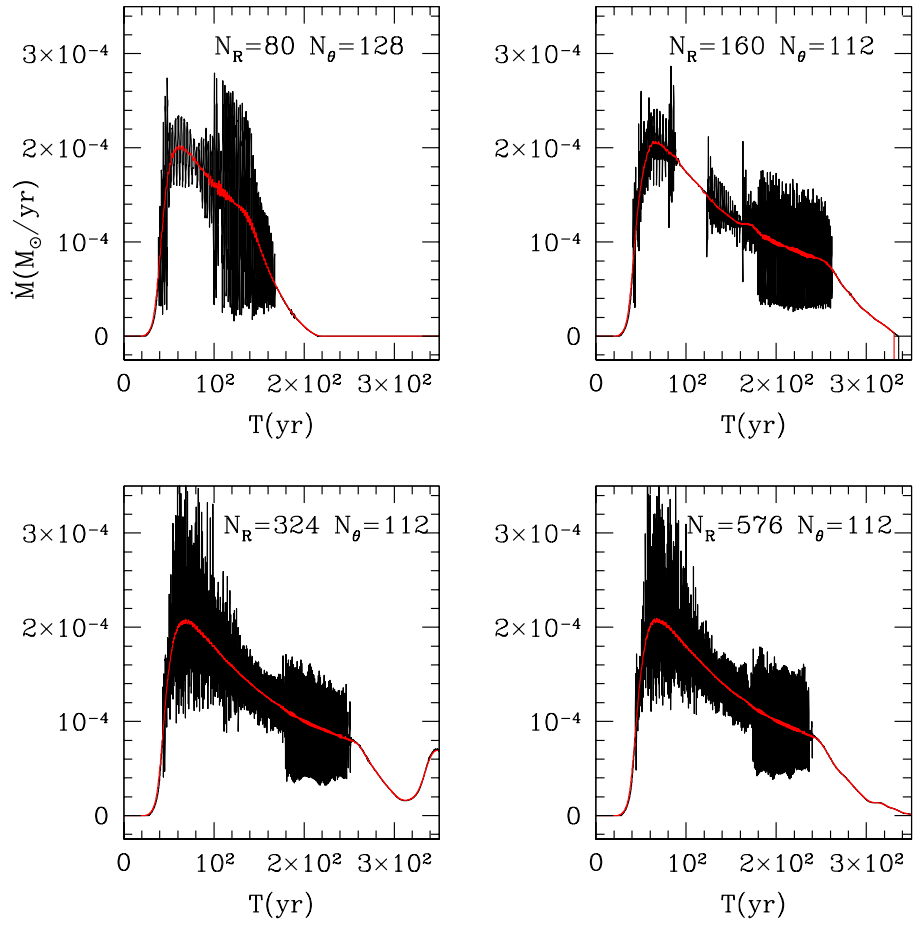} \caption{ The disk's mass
accretion rate through the innermost boundary with time for models
with four different resolutions. The red lines are the mass
accretion rates after being smoothed. }\label{fig:reso}
\end{figure}

Figure \ref{fig:reso} shows the mass accretion rates during an
outburst for models with four different numerical resolutions.  In
general the vertical linear resolution must be higher than the
radial linear resolution to resolve the vertical structure of the
disk, particularly the exponential density drop in the disk
photosphere. With $N_{R}=320$ grid cells and $N_{\theta}=112$ grid
cells, every grid is 5 times longer radially than vertically.

The simulations have a similar maximum mass accretion rate and outburst
timescale as long as $N_R \gtrsim 100$ grid cells.  This suggests that
the models are, crudely speaking, converged.

Notice that models with N$_{R} \ge 324$ grid cells have complex
substructure during outburst.  This is caused by convection in the
TI high state region.  Lower resolution models barely resolve the
convective eddies and thus damp away this substructure.

\subsection{Boundary effects}

\begin{figure}
\epsscale{.80} \plotone{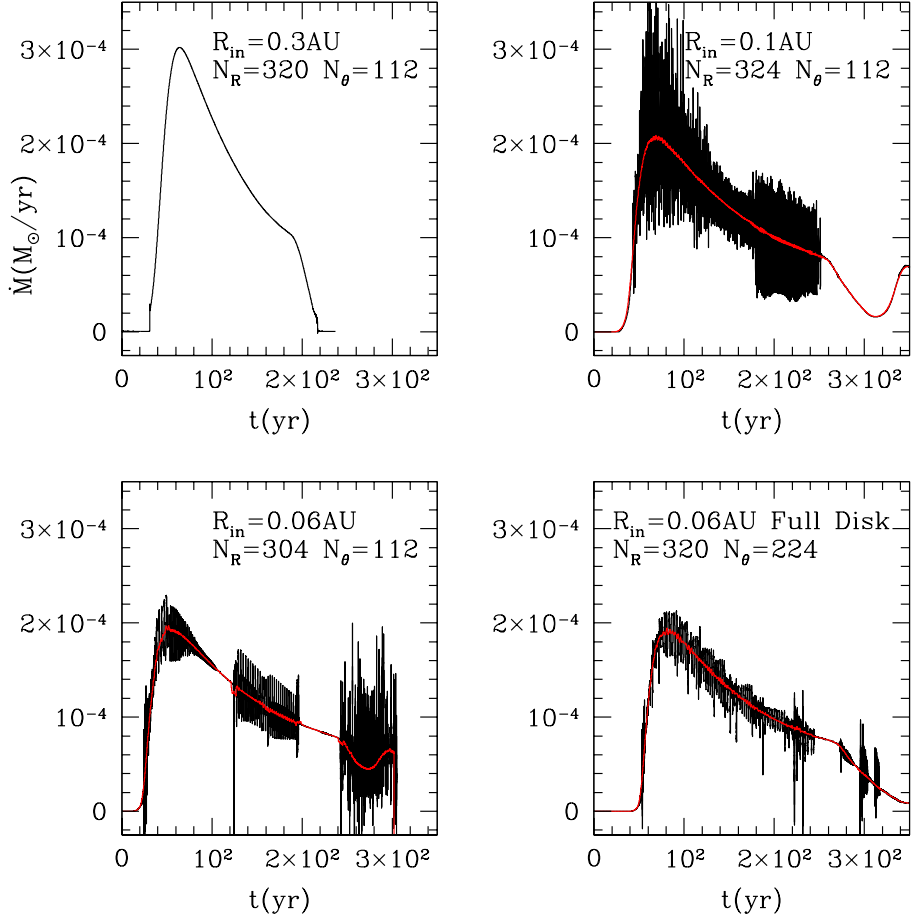} \caption{ The first three panels
show the disk's mass accretion rate for half-disk models with
different inner radius (0.3 AU to 0.06 AU). The lower right panel
shows the disk's mass accretion rate for a full disk with the inner
radius 0.06 AU.  The red lines are the mass accretion rates after
being smoothed.}\label{fig:diffsetdm}
\end{figure}

We have also explored the effect of varying the inner boundary
radius $R_{in}$ on the models.  Figure \ref{fig:diffsetdm} shows
$\dot{M}(t)$ for a set of models with $0.06\au \le R_{in} \le
0.3\au$, with one two-sided (not equatorially symmetric) model at
$R_{in} = 0.06\au$.  All models have $N_R = 320$ and $N_\theta =
112$, except the two-sided model, which has $N_\theta = 224$.

The outburst amplitude and duration are nearly independent of
$R_{in}$ as long as $R_{in}$ is far enough in that the TI can be
triggered; the TI is not triggered when $R_{in} = 0.3\au$, and so
the outburst is slightly stronger and shorter.  One implication of
the lack of sensitivity of the outburst profile to $R_{in}$ is that,
while the TI can sharpen the outburst slightly, the outburst
timescale and maximum mass are largely set in the outer disk: where
the MRI becomes active, how much mass is accumulated during the
quiescent stage, and so on.

When the TI is triggered, $\dot{M}(t)$ exhibits strong short-timescale
variations (panel (b,c,d) of Fig. \ref{fig:diffsetdm}).  This is caused
by convection during the transition between the TI high and low
states; the large temperature gradients at the transition make it highly
convectively unstable by the Schwarzschild criterion (see \S 4 for
details).

Convection at the high state-low state transition radius changes the
character of the $\dot{M}(t)$ variability in 2D models compared to 1D
(vertically averaged) models.  In 1D evolution models flickering at any
radius will propagate through the entire inner disk, making the entire
inner disk flicker between the low and high states.  Thus the mass
accretion rate varies violently during 1D evolutions.  To avoid this,
Bell \& Lin (1994) used an $\alpha$ prescription designed to make
$\alpha$ increase sharply from the low to the high states.  In 2D
models, on the other hand, flickering propagates both radially and
vertically.  For example, consider a region near the midplane that wants
to jump to the high state will soon experience convective cooling and
return to the low state.  This acts to localize and dampen low
state-high state flickering.

For models with $R_{in} = 0.06 \au$ (panel (c) in Fig.
\ref{fig:diffsetdm}), $\dot{M}(t)$ varies less violently than in
models with $R_{in} = 0.1\au$ because the inner edge is far from the
convection associated with the low state high state transition.  As
the resolution increases, however, the $R_{in} = 0.06\au$ model
becomes more and more variable.  These models, and especially the
convection zones, are not converged even at our highest resolution.
This is not a major concern, as the fine structure of convection
in our axisymmetric, phenomenological  $\alpha$ viscosity model will
not be the same as the fine structure of convection in a three
dimensional, MHD turbulent disk; further resolving this structure
would not improve the model.

One aspect of the details of convection is worth commenting on.
Panel (d) in Figure \ref{fig:diffsetdm} shows a full-disk model,
which exhibit less variability than the half-disk models. Convection
in the full-disk models can penetrate, rather than reflecting back
from, the midplane.  This feature of disk convection is consistent
with linear theory (Ruden et al. 1988), which shows that the most
unstable mode has no nodes in $v_z$.  Because the maximum mass
accretion rates and the outburst duration times are similar for
half-disk and full-disk models, most of our simulations have been
carried with one-sided models to save computation time. However, we
need to use the full-disk models to study convection during the
outburst.

\section{Vertical Structure and Convection}

Though our 2D simulations exhibit complex time-dependent behavior, the
vertical structure of the simulated disk can be well fit by a simple
analytical calculation plus some basic assumptions.

Assuming local energy dissipation
and with a given effective temperature and viscosity parameter
$\alpha$, the disk structure can be defined by the following
equations. Assuming all the viscosity-generated energy radiates
vertically, we have
\begin{equation}
\frac{9}{4}\nu_{V}\rho\Omega^{2}=\frac{d}{d z}\left( F\right) \, ,
\label{eq:rt}
\end{equation}
where the vertical flux $F$ is
\begin{equation}
F=\frac{4\sigma}{3}\frac{dT_{\tau}^{4}}{d{\tau}}\,, \label{eq:flux}
\end{equation}
the viscosity is
\begin{equation}
\nu_{V}=\alpha c_s^2/\Omega \,, \label{eq:nu}
\end{equation}
and
\begin{equation}
d\tau=-\rho \kappa(T,p) dz \,\, . \label{eq:tau}
\end{equation}
With hydrostatic equilibrium,
\begin{equation}
\frac{dp}{dz}=-\frac{GM_{*}\rho z}{R^{3}}  \, , \label{eq:hydro}
\end{equation}
and the equation of state,
\begin{equation}
p=\frac{k_{B}}{\mu}\rho T \, . \label{eq:EOS} \label{eq:eos}
\end{equation}
$\rho$,$T$, and $p$ are functions of Z. Equation \ref{eq:rt} can be
integrated towards the midplane given $T_{eff} = T(\tau=2/3)$ as the
boundary condition. The integration is halted at $z_f$, where the
total viscous heating is equal to $\sigma T_{eff}^4$. In general
$z_f \ne 0$ so we change the initial conditions and repeat the
integration until $z_f = 0$, in which case we have the
self-consistent vertical structure solution. This procedure is
identical to that described in Zhu et al.  (2009); we will call
these local vertical multi-zone (LVMZ) models.

Figure \ref{fig:2} shows the vertical structure at 0.1 AU from the
2D simulation and that from the LVMZ with the same central
temperature. Though the 2D model cannot resolve the region where
$T\sim$ $10^{4} K$, the 2D model otherwise agrees with the LVMZ very
well, which means that at 0.1 AU the local treatment is a good
approximation even during the TI high state, and that radial energy
transport is not important. We find that in general the simulation
follows the 'S' curve calculated by the LVMZ very well during the
whole TI activation process (see Appendix B).

In the lower left panel of Figure \ref{fig:2}, the pressure plateau
around $\tau\sim 10$ is caused by the rapid increase in opacity with
increasing temperature near $\sim 10^{4} K$ as H ionizes (see the
appendix in Zhu et al. 2009).  This means that the optical depth can
change significantly while the pressure remains nearly the same.

The large vertical temperature gradient causes this region to become
convectively unstable. As shown in the lower right panel, the dense
gas is on top of the thin gas at $\tau<100$ which is clearly
unstable against convection or Rayleigh-Taylor instability. Though
this convective region is only at the surface and barely resolved at
0.1 AU, it extends to the whole disk at 0.2 AU, where convective
patterns even penetrate the mid-plane (Fig. \ref{fig:convection}).

Convection occurs wherever the Schwarzschild criterion is violated:
$\nabla_{S} \equiv d\log T/d\log P
> (\gamma-1)/\gamma = 0.4$ in our case ($\gamma=5/3$).
For an optically thick disk, if the opacity can be described as
$\kappa=\kappa_{0}T^{\beta}P^{\epsilon}$ \citep{bell94,zhu2009},
\cite{rafikov2007} derives $\nabla_{S}>$(1+$\epsilon$)/(4-$\beta$).
Thus, the disk is convectively unstable when its temperature is in
the range with the opacity satisfying
(1+$\epsilon$)/(4-$\beta$)>($\gamma-1$)/$\gamma$=0.4 .

Before the outburst, the disk is convectively stable. The active
layer has a lower temperature than the dust sublimation temperature.
Thus, $(1+\epsilon)/(4-\beta) \simeq 0.3$, using the Bell \& Lin
(1994) opacities. The numerical simulation also shows $\nabla_{S}$
is slightly smaller than ($\gamma-1$)/$\gamma$=0.4, which implies it
is convectively stable. \footnote{We assume $\gamma=5/3$ in our
simulations.  But at temperatures somewhat lower than the H$_{2}$
dissociation temperature, $\gamma\sim$ 7/5, which implies
($\gamma-1$)/$\gamma\sim$0.3. Thus, the dusty disk may be
convectively unstable. A self-consistent thermodynamic treatment
will be considered in a later paper.} The dead zone is isothermal
without any energy generation, which means it is also convectively
stable with $\nabla_{S}\sim$0.

During the TI high state, however, the disk exhibits a variety of
convective features at $R<0.6 \au$, where $T > 2000\K$.  The inner disk
consists of three distinct convective regions:

1) At $0.35 \au < R < 0.6 \au$, the disk is hot enough that the dust
has sublimated and the gas is in molecular form ($2000 \K < T < 5000
\K$) with TiO and H$_{2}$O opacities dominating the opacity. Based
on the opacity given by Bell \& Lin (1994),
$(1+\epsilon)/(4-\beta)\sim$1 so that the disk is convectively
unstable.  Our numerical simulations have also shown that
$\nabla_{S}>0.4$ in these regions (changing from 0.4 at $R \sim0.6
\au$ to 1 at $R\sim 1 \au$). However, convection turns out to be
inefficient in these regions, with convective speeds far less than
the sound speeds (lower right panel of Fig. \ref{fig:convection}).

2) At 0.15 AU < R < 0.35 AU, the midplane is hot enough that
hydrogen begins to be ionized. The midplane is thermally unstable, and
$(1+\epsilon)/(4-\beta)$ even becomes negative. The disk heats up
until the midplane temperature $> 2 \times 10^4\K$ and it becomes
thermally stable.  The surface, however, is still far cooler than
$2\times 10^{4}\K$, with the effective temperature $\sim 2000 \K$.
At some height in the disk there is a transition where the opacity
depends steeply on the temperature and convection is strong.  At
these radii the transition is close to the midplane and strong
convection stirs the entire disk.  The largest scale eddies
penetrate the midplane.

Figure \ref{fig:trace} shows a velocity map of the convective
eddies in region (2).  We study them by inserting trace particles.
These particles rise up with the hot 'red' fingers which are in the
TI high state. When they approach the surface, they enter the region
which is cool, dense and in the TI low state. Then they fall with
the cool 'green' fingers to the midplane where they will be heated
up again. The convective velocity can be estimated by
assuming the particles in the red fingers are accelerated by the
buoyancy force from the surrounding dense green fingers. Consider
the red fingers are in TI high state with T$\sim$20,000 K and
$\rho_{r}$, while the green fingers are in TI low state with
T$\sim$2,000 K and $\rho_{g}\sim10\rho_{r}$. Thus, in the rest frame
of the green fingers
\begin{equation}
v_{r}\sim\left(\frac{H_{e}a}{2}\right)^{1/2}\sim\left(\frac{H_{e}
g_{h}}{2}\frac{\rho_{g}-\rho_{r}}{\rho_{r}}\right)^{1/2}\sim\left(4.5
H_{e} g_{h}\right)^{1/2}
\end{equation}
where $H_{e}$ is height of the eddy.
$g_{h}$=GM$_{*}$/R$^{3}$$\times$H/2. While the velocity of the red
fingers in the rest frame of the green fingers is
\begin{equation}
v_{g}\sim\left(\frac{H_{e}
g_{h}}{2}\frac{\rho_{g}-\rho_{r}}{\rho_{g}}\right)\sim\left(0.5 H
g_{h}\right)\,\label{eq:vturb}
\end{equation}
The real eddy velocity should between v$_{g}$ and v$_{r}$. If
H/R$\sim$0.1, $H_{e}$=0.5 H and R=0.2 AU, we derive v$_{g}\sim$2.3
km s$^{-1}$ and v$_{r}\sim$ 7.1 km s$^{-1}$. Thus the average eddy
velocity should be around 4 km s$^{-1}$. The maximum velocity should
not be larger than twice the average velocity. The lower left
panel shows the velocity of these eddies, whose velocity is around 4
km s$^{-1}$. The maximum velocity takes place at the surface.
Although convection is subsonic at the midplane, it is
supersonic at the surface (lower right panel in Fig.
\ref{fig:convection}).

3) At $R<0.15 \au$ most of the region is already in the TI high
state with $T>2\times 10^{4} \K$. Thus $(1+\gamma)/(4-\beta)\sim
0.3$, which is similar to the numerical simulation. The surface
region where the high state joins to the low state is, however,
still convectively unstable (refer to Fig.  \ref{fig:2} which shows
$\nabla_{S}$ is very large at the transition).  Unfortunately our
simulation barely resolves this region.

Generally, convection in our simulation is strongly correlated with TI
(which is not surprising, since both are related to steep gradients in
the opacity).  In the outer region where the TI is inactive, convection
is weak and inefficient. When the TI is active, the convection patterns
depend on the vertical height of the point where hydrogen begins to
ionize.  If this transition is close to the midplane, convection is
strong and penetrates the midplane.  Otherwise, only the surface is
affected by convection.

\section{Discussion}

Using our adopted parameters, our 2D simulations produce outbursts which
are similar to FU Orionis outbursts, with comparable maximum mass accretion
rates ($\sim 2 \times 10 ^{-4}\msunyr$) and decay timescales ($\sim
100 $yrs). The outbursts begin with the GI transferring mass to the
inner disk until it is hot enough to activate the MRI; at that point
the disk accretes at a much higher mass accretion rate because of the
much higher efficiency of the MRI in transferring angular momentum.

Our outburst scenario is similar to that of \cite{armitage01}. However,
their outbursts were at a much lower mass accretion rate,
$\sim 10^{-5}\msunyr$, and lasted a longer time, $\sim 10^{4}$ yr.  As
discovered by Book \& Hartmann (2001) and explained in Zhu et al.
(2009), their weaker and slower outbursts are due to their lower MRI trigger
temperatures ($T_{M}\sim 800 \K$) and smaller $\alpha_{M}=0.01$.  Specifically,
with a lower $T_{M}$, the MRI is triggered at a larger
radius with a lower surface density, resulting in less mass accumulation
in the inner disk.  Combined with a smaller $\alpha$, this results in
a lower mass loss rate and a longer viscous timescale.

Our choice of $T_{M}\sim 1200 \K$ is close to the temperature range
at which dust is expected to sublimate.  This suggests that the
mechanism for turn-on of the MRI thermally may be controlled by the
elimination of the small dust grains that effectively absorb
electrons and thus quench the MRI.

Our radiative transfer modeling of FU Ori (Zhu \etal 2007) suggested
that to fit the {\em Spitzer Space Telescope} IRS spectrum, the
rapidly-accreting, hot inner disk must extend out to $\sim 1$~AU,
inconsistent with a pure thermal instability model \citep{bell94}.
In our 2D results we find MRI triggering at about 2 AU, much more
consistent with the {\em Spitzer} data and our previous
analysis in Zhu et al. 2009.  As
FU Ori has a smaller estimated central star mass (0.3 M$_{\odot}$)
than the 1 M$_{\odot}$ used in our simulations, its hot inner disk should be
slightly smaller than derived here, but still in agreement
with observations.

Non-Keplerian rotation has not been observed from optical to near-IR
(5$\mu$m) lines \citep{zhu2009b} since these lines come from inner
disk within 1 AU where the disk is not massive. However, due to the
fact that the disk is massive and gravitationally unstable at large
radii (>2 AU), the disk has slightly sub-Keplerian rotation there and
this effect may be observable in the far-IR or submm \citep{lb2003}.

We cannot calculate meaningful optical (B magnitude) light curves
for comparison with the historical observations of FU Ori objects
because our inner disk radius is too large; this results in our
maximum effective temperatures being too low in comparison with
observations. (This constraint was due to the very long computation
times required during outburst; moving the inner radius inward
greatly lengthens the computing time.)  We therefore computed a
1$\mu$m light curve as a prediction to compare with future
observations (Fig. \ref{fig:lc2}), assuming that the disk radiates like
a blackbody at each radius's effective temperature; the lower panel
shows the disk's surface and central temperature at 0.06 AU.
The outburst has rise time $\sim$ 10 years, which is
similar to V 1515 Cyg but longer than FU Ori. The thermal
instability is triggered on a much shorter timescale: as the lower
panel shows, the disk's central temperature rises sharply within 1
year. However, since the mass at the inner disk is still low, the
mass accretion rate and the effective temperature cannot increase
significantly until the MRI transfers more mass from the outer disk
to the inner disk on a viscous timescale. Overall, our simulation
agrees with V 1515 Cyg reasonably well but has a longer rise time
than FU Ori and V 1057 Cyg.  This might be improved by using a
smaller inner radius, a consistent central mass, and a
self-consistent boundary condition \citep{kley1999}.

Our simulation suggests that the short-timescale variations
appearing in the light curves of FU Ori \citep{kenyon2000} might be
caused by convection in the high state to low state transition
region.  The observations show $\Delta M \simeq 0.035$ and
timescales of less than one day.  The amplitude of the variability
in the 2D models is $\Delta M \sim 0.2$ (see Fig. \ref{fig:lc2}). In
3D the amplitude of the variability will be reduced because there
will be several convective eddies at each radius and averaging over
these will reduce the variability.  If we assume that the azimuthal
extent of a convective eddy is $\sim H$, then there will be $m = 2
\pi R/H$ convective eddies and the variability will be reduced by a
factor of $m^{1/2} \sim 5$ if $H/R \sim 0.2$, suggesting a
variability of $\Delta M \sim 0.04$.  This agreement the data is
encouraging, but it is likely that the details of convection in a
magnetized, three dimensional disk will be different from that found
in our viscous, axisymmetric models.  The modest success of our
model, however, might motivate the development of energetically
self-consistent 3D magnetohydrodynamical (MHD) models of the inner
disk of FU Ori.

The timescale of variability in our 2D models (months) is much
longer than the observed 1 day.  The timescale of variability might
be shortened by (1) increased resolution, since the timescale of
variability in our models increases with resolution; (2) switching
off all components of stress in Equation \ref{eq:stress}
except the r-$\Phi$ component, since the other components act to
retard and smooth convection (3) replacing the
viscosity model used here by MHD turbulence, since the viscosity
tends to suppress small scale structure; (4) 3D effects, since one
might expect to see variability on timescales of $2\pi/(m\Omega)
\sim 1$ day, using $R = 0.2 \au$ and our estimate for $m$ from
above. Overall, although convective behavior revealed by our
current simulations are very preliminary (they depend on the
simulation resolution, opacity, details of the turbulence etc.), it
seems reasonable that convection accounts for the short-timescale
variability of FU Orionis objects.

Hartmann, Hinkle, \& Calvet (2004) found that some mildly supersonic
turbulence ($\sim$ 2 c$_{s}$) was needed to explain the width of
$^{12}$CO first-overtone lines of FU Ori .  High resolution spectral
modeling suggests that CO lines broadened with $\sim$4 km/s turbulent
velocity fit the observations well (Zhu et al. 2007). Hartmann, Hinkle,
\& Calvet (2004) proposed that the MRI could be a mechanism to drive the
supersonic turbulence, while here we suggest that convection could
also drive the turbulence.  Convection is subsonic at the midplane in
the TI high state, but as the convective eddies travel to the surface
the temperature drops and convection becomes supersonic (see the
lower right panel of Fig. \ref{fig:convection}).  Numerical simulations
(lower left panel of Fig. \ref{fig:convection}) and the simple
analytical estimate (Eq. \ref{eq:vturb}) suggest the 5 km s$^{-1}$ is a
typical convective velocity.

We also studied the disk central temperature at the stage when the
MRI is activated and then moves inwards. The transition between the
TI active and inactive region is sharp and occurs over a radial
interval comparable to the disk scale height. The fronts travel at
2.7 $\kms$, which is almost the sound speed at T = 1000 K. This
thermal front velocity is much higher than $\alpha c_{s}$ predicted
by Lin et al. (1985) and we will discuss this phenomenon in detail
in our next parameter study paper. Notice that in our model the
stress is assumed to respond instantaneously to a change in the
temperature caused by the radial energy diffusion ; this may indeed
be the case if magnetic field mixes into the newly heated gas from
the hot side of the transition front. If the magnetic field takes
longer than $\sim \Omega^{-1}$ to build up then the front will travel
more slowly.

Radiation pressure starts to become significant during the TI high
state, with P$_{r}$/P$_{t}\sim$ 0.3 in the innermost regions. We
have conducted some experiments including radiation pressure using
the diffusion approximation and find generally similar behavior to
what has been shown in this paper, though the disk is slightly
thicker with the inclusion of radiation pressure.  Again, we note
that our inner radius is considerably larger than those of T Tauri
stars, so radiation pressure might be more important as accreting
material joins the central star.
\section{Conclusions}

We have studied the time evolution of FU Orionis
outbursts in a layered disk scenario, using an axisymmetric two
dimensional viscous hydrodynamics code with radiative transfer.  Both
the MRI and GI are considered as angular momentum transport mechanisms
and are modeled by a phenomenological $\alpha$ prescription. Besides the
surface of the disk ionized by cosmic rays and X-rays, the MRI is also
considered to be active as long as the disk's temperature is above a
critical temperature $T_{M}$, so that thermal ionization can provide
enough ions to couple the plasma to the magnetic field.

Starting with a massive disk equivalent to a constant $\alpha$ disk
with $\dot{M}$=10$^{-4}$ M$_{\odot}$/yr and $\alpha$=0.1, several
outbursts appear during our simulation. These outbursts begin with
the GI transferring mass to the inner disk until the inner disk
gradually becomes gravitationally unstable. Eventually the disk at 2
AU is hot enough to trigger the MRI and then the MRI active region
expands throughout the whole inner disk.  With the increasing
temperature since the MRI is transferring more mass to the inner
disk, the TI is triggered and a ``high state'' expands outwards to
0.3 AU. The outburst state lasts for hundreds of years and drains
the inner disk; the disk then returns to the low state.

Convection appears at the inner disk during the TI high state. In
the outer region ($0.35<R<0.6 \au$), where the TI is inactive,
convection is weak and inefficient.  At 0.15<R<0.35 $\au$ during
outburst, there is a hydrogen ionization zone at $z\sim H$, and
strong convection is present.  The convection even penetrates the
midplane to form antisymmetric convective rolls.  In the innermost
region most of the disk is hot, the hydrogen ionization zone is
close to the photosphere, and weak convection is confined to the
disk surface.

Our model shows maximum accretion rates and outburst duration
timescales similar to those of the known FU Orionis objects. Unlike
the pure TI theory, our model produces an AU scale inner disk with
high mass accretion rate, in agreement with the observational
constraints.

With some simple assumptions our simulations have produced FU
Orionis-like outbursts.  They also pose some interesting questions
for future study.  In a later paper we will compare our 2-D results
to 1-D simulations; the latter will allow us to consider longer disk
evolutionary timescales.

A dynamical GI simulation at AU scale, with both self-gravity and
radiation, will be important to understand angular momentum
transport by GI and the thermal structure of the inner disk.
However, considering the large Keplerian velocity of the inner disk,
a self-consistent radiation hydrodynamic scheme will be needed.  Our
simulations have also shown that convection is important in FU
Orionis objects due to the large vertical temperature gradient at
the hydrogen ionization zone. A fully 3-D simulation with both MHD
and radiation, together with a self-consistent treatment of the
equation of state, will be essential to understanding the role of
convection. Also the very inner disk where the MRI is thermally
activated by the irradiation and the boundary layer may also play a
role in the disk evolution.

Z. Zhu wants to thank Dr. R. Durisen and K. Cai for the inspiring
discussion about their radiative transport scheme. This work was
supported in part by NASA grant NNX08A139G, by the University of
Michigan, and by a Sony Faculty Fellowship, a Richard and Margaret
Romano Professorial Scholarship, and a University Scholar
appointment to CG. JCM was supported by NASA's Chandra Fellowship
PF7-80048.

\section{Appendix A: Radiative cooling scheme tests}

To test the radiative transfer scheme (and particularly the boundary
condition at the disk photosphere), we have constructed a disk model
in which the viscous heating is turned off when $\tau<100$.  This
makes the vertical flux constant for $\tau < 100$ and allows us to
test our scheme against a gray atmosphere model. The disk structure
shown here is that calculated with Method 1 as discussed in \S 2.2.
The structure calculated with Method 2 is even in a better agreement
than Method 1, since it directly uses the gray atmosphere structure
as the boundary condition.

The left panel in Figure \ref{fig:1} shows the flux through the
atmosphere at different heights for one radius, while the dotted
line is $F=\sigma T_{eff}^{4}$ with T$_{eff}$ derived from the
simulation. The right panel shows the disk temperature structure at
this radius from the 2D simulation (solid line) and the theoretical
gray atmosphere temperature structure (T$^{4}$=3/4
T$_{eff}^{4}$($\tau$+2/3) (dotted line).  The vertical structure in
the 2D simulation agrees well with the gray atmosphere structure,
and improves on the original scheme of \cite{Cai2008} by eliminating
a temperature jump at the photosphere.

\section{Appendix B: Thermal instability}

The thermal instability (TI) model was developed to account for
outbursts in dwarf novae systems (Faulkner, Lin, \& Papaloizou 1983).
But it also has features that make it attractive for explaining FU Ori
outbursts \citep{bell94}. Although in this paper we find that MRI
triggering is mainly responsible for initiating FU Orionis outbursts,
the TI still exists and can be triggered during the outburst.  With the
increasing mass accretion rate by the MRI operating, the innermost disk
is heated up to hydrogen ionization and the disk becomes thermally
unstable.  Finally, TI sets in and the innermost disk enter a high
(fully ionized) state.

The basic idea behind the TI can be illustrated by the disk
$\Sigma-T_{c}$ relationship ('S' curve, Fig.\ref{fig:scurve}). The disk
is in thermal equilibrium on this curve.  To the left of the curve
cooling exceeds heating and to the right heating exceeds cooling.  As
the disk central temperature reaches $3000 \K$ the disk transitions from
the low-state lower branch to the high-state ($T_c \sim 2 \times 10^4
\K$) upper branch; it cannot sit on the unstable middle branch.

\subsection{'S' curve of 2D models}

As discussed in \S 4.1, the LVMZ method \citep{zhu2009} produces
similar vertical structure to our 2D models.  Here, we use the LVMZ
to calculate the disk's $\Sigma-T_{c}$ relation at 0.1 AU in Figure
\ref{fig:scurve}. The solid line is the equilibrium curve, while the
crosses show $\Sigma$ and $T_{c}$ at 0.1 AU during 2D simulations;
the disk moves from the lower left to the upper right with time. The
crosses trace the LVMZ curve remarkably well.  Since the LVMZ
assumes the local cooling through surface balances local viscous
heating, it follows that 2D effects, such as radial transport of
heat, are not important in this region.

\subsection{'S' curve of 1D models}

In one zone model, where the midplane opacity is used at all $z$ and the
total surface density replaces $\rho(z)$, the 'S' curve can be obtained
via
\begin{equation} F_{vis}=F_{rad} \,.
\end{equation}
Thus,
\begin{equation}
\frac{9}{4}\alpha\Omega\Sigma c_{s}^{2}=2\sigma T_{eff}^{4}\,,
\end{equation}
where $\Sigma$ is the surface density and $c_{s}$ is the sound speed
at the local central disk. For an optically-thick disk, we have
T$_{c}^{4}$=$\frac{3}{8}$$\tau_{R}$T$_{eff}^{4}$, where
$\tau_{R}$=$\kappa_{R}$$\Sigma$ and $\kappa_{R}$ is the Rosseland
mean opacity \citep{Hubeny1990}.  Then,
\begin{equation}
T_{c}^{3}=\frac{27}{64\mu\sigma}\alpha\Omega\Sigma^{2} {\cal{R}}_{c}
\kappa_{R}
 \,,
\end{equation}
where $\cal{R}$$_{c}$ is the gas constant, and $\mu$ is the mean
molecular weight .  This $T_{c}$ and $\Sigma$ relationship is shown
in Figure \ref{fig:scurve}, which differs significantly from the
LVMZ 'S' curve.

These 'S' curves again make the point that the observed large outer
radius, 0.5 $-$ 1 AU, of the high accretion rate portion of the FU
Ori disk poses difficulties for the pure thermal instability model.
Even in the situation explored by \cite{Lodato2004}, in which the
thermal instability is triggered by a massive planet in the inner
disk, the outer radius of the high state is still the same as in
\cite{bell94}, $\sim$ 20 $\rsun$. The relatively high temperatures
($2-4 \times 10^3$~K) required to trigger the thermal instability
(due to the ionization of hydrogen) are difficult to achieve at
large radii; they require large surface densities which in turn
produce large optical depths, trapping the radiation internally and
making the central temperature much higher than the surface
(effective) temperature.

\clearpage

\clearpage

\begin{figure}
\epsscale{.80} \plotone{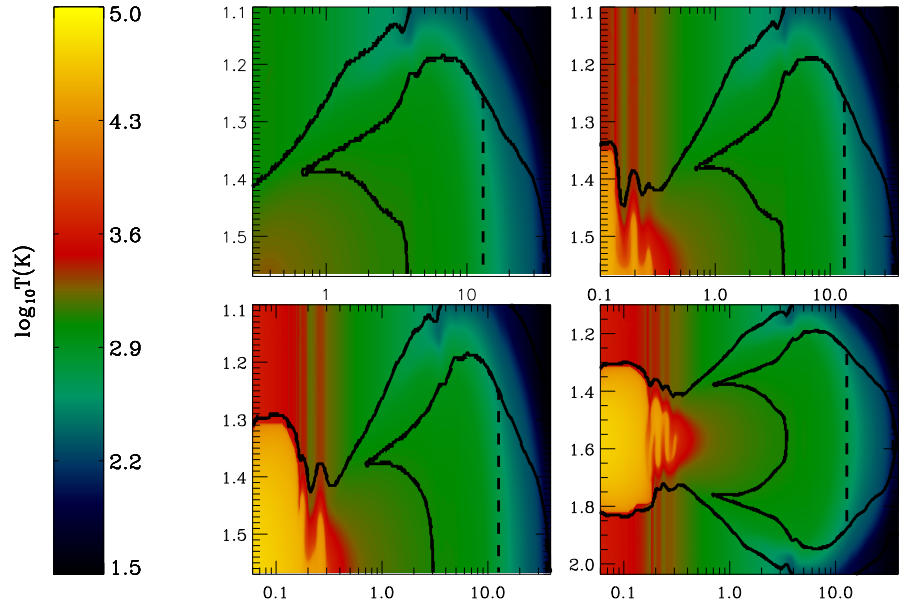} \caption{ The temperature contours
at the stage when their mass accretion rates reach maximum for the
four models shown in Figure \ref{fig:diffsetdm}.
 }\label{fig:diffset}
\end{figure}

%\begin{figure}
%\epsscale{.80} \plotone{pre.eps} \caption{ }\label{fig:pre}
%\end{figure}

%\begin{figure}
%\epsscale{.80} \plotone{titrig.eps} \caption{ }\label{fig:titrig}
%\end{figure}

%\begin{figure}
%\epsscale{.80} \plotone{tihigh.eps} \caption{ }\label{fig:tihigh}
%\end{figure}

%\begin{figure}
%\epsscale{.80} \plotone{tolow.eps} \caption{ }\label{fig:tolow}
%\end{figure}

%\begin{figure}
%\epsscale{.80} \plotone{relow.eps} \caption{ }\label{fig:relow}
%\end{figure}

\clearpage

\begin{figure}
\epsscale{.80} \plotone{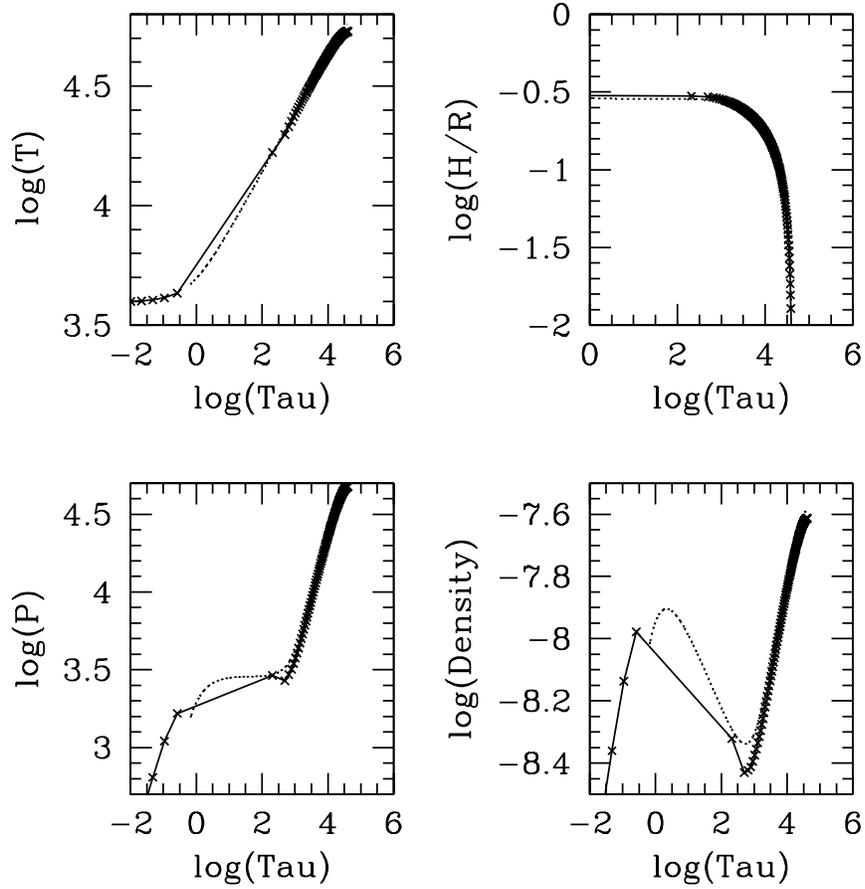} \caption{ The disk's vertical
structure at 0.1 AU in the TI active stage. The crosses are from the
2D simulation grids, while the dotted lines are the disk
structure at the same central temperature calculated by the
LVMZ method. }\label{fig:2}
\end{figure}

\clearpage

\begin{figure}
\epsscale{.80} \plotone{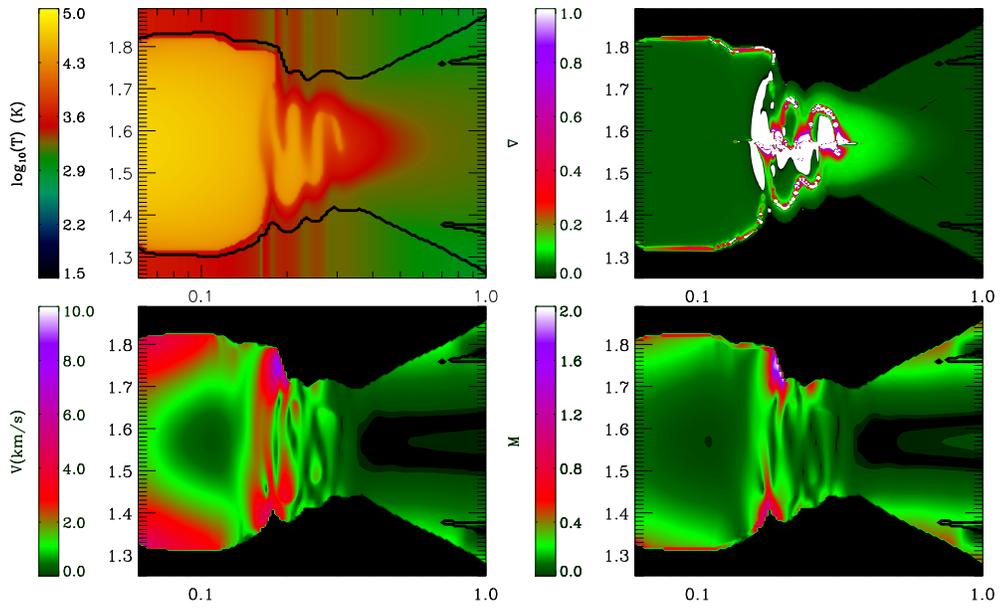} \caption{The contours for the
temperature, temperature gradient, velocity and Mach number at the
TI active stage. The white regions in the temperature gradient
contour are due to the gradient larger than the shown range from 0
to 1. The optically thin region has been masked off in order to have
the convective patterns in the optically thick region stand out.
}\label{fig:convection}
\end{figure}

\begin{figure}
\epsscale{.80} \plotone{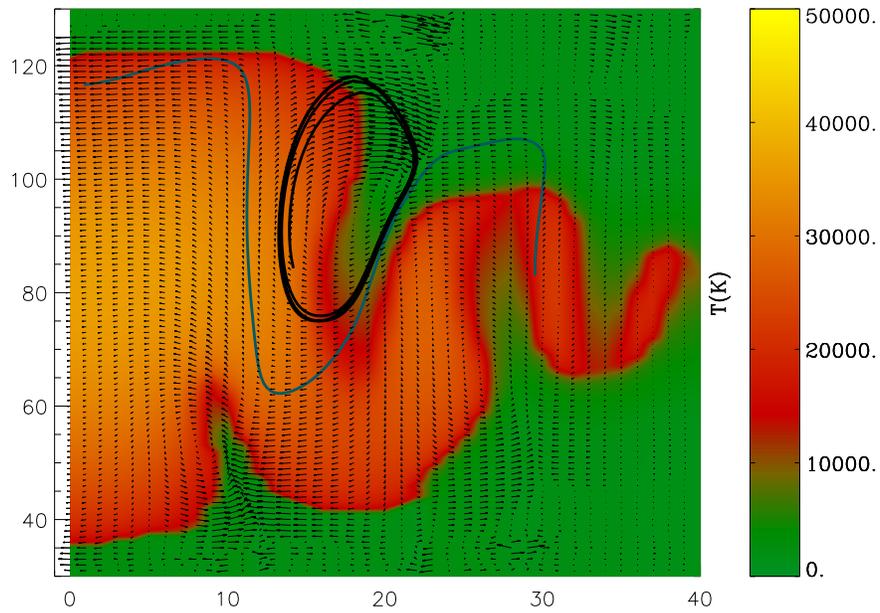} \caption{The velocity vectors of
the computational grid are plotted on top of the temperature
contours at the TI active region. The black and blue curves are the
traces of two test particles which are originally placed at the
midplane at 0.4 and 0.8 AU.}\label{fig:trace}
\end{figure}

%\begin{figure}
%\epsscale{.80} \plotone{uplow.eps} \caption{ }\label{fig:uplow}
%\end{figure}
\clearpage

\begin{figure}
\epsscale{.80} \plotone{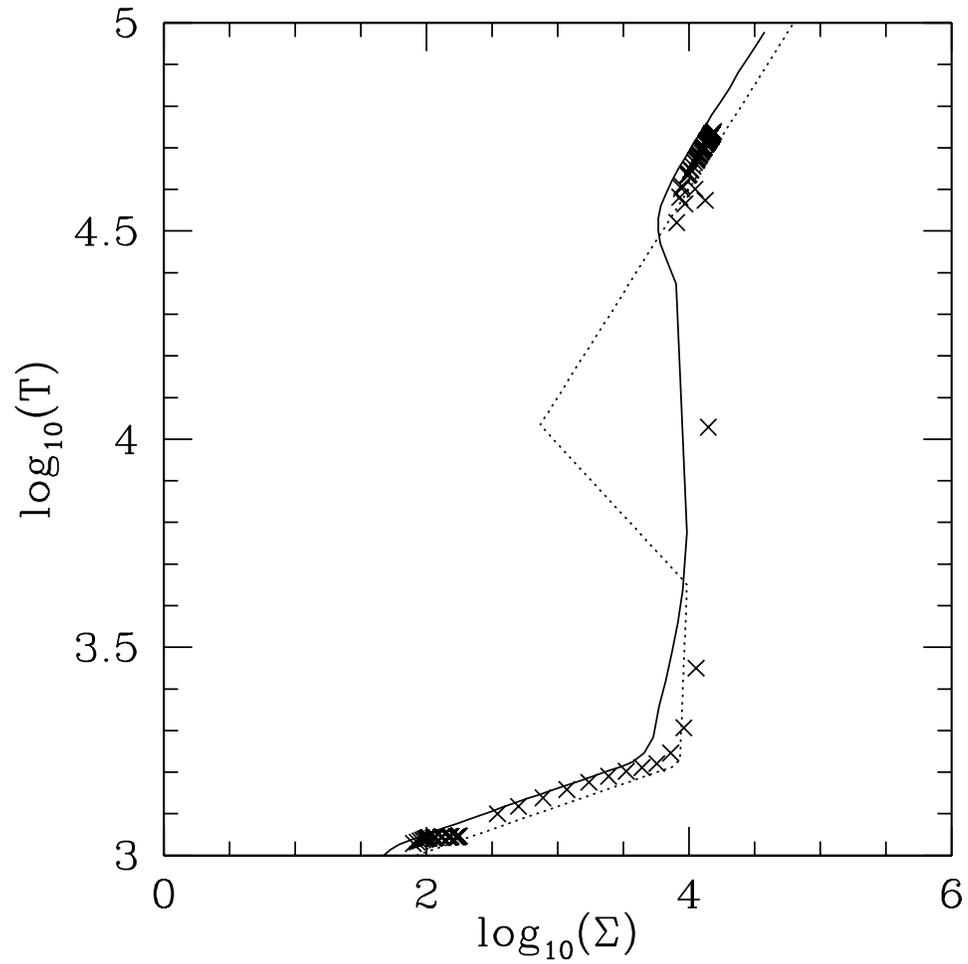} \caption{The 'S' curve calculated
by the LVMZ (solid curve) and the midplane approximation (dotted
curve) at 0.1 AU. The crosses trace the $\Sigma$ and T$_{c}$ from
the 2D simulation. }\label{fig:scurve}
\end{figure}

%\begin{figure}
%\epsscale{.80} \plotone{scurvecompare.eps} \caption{
%}\label{fig:scurvecompare}
%\end{figure}

\begin{figure}
\epsscale{.80} \plotone{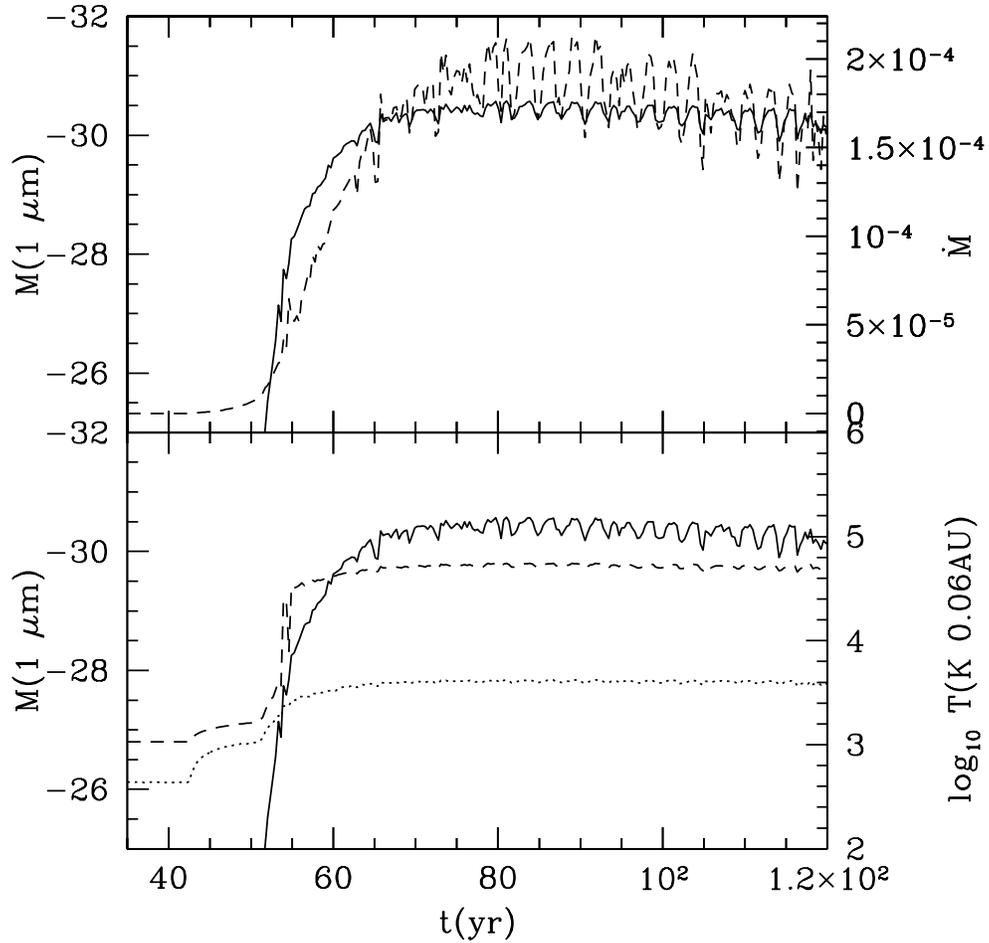} \caption{ The solid curve shows
the 1$\mu$m magnitude derived from the 2D disk at the beginning of
the outburst. The dashed curve in the upper panel shows the mass
accretion rate, while the dashed and dotted curves in the lower
panel show the midplane and surface temperature at 0.06 AU
respectively. }\label{fig:lc2}
\end{figure}

%\begin{figure}
%\epsscale{.80} \plotone{figure6.ps} \caption{ }\label{fig:6panel}
%\end{figure}

\clearpage

\begin{figure}
\epsscale{.80} \plotone{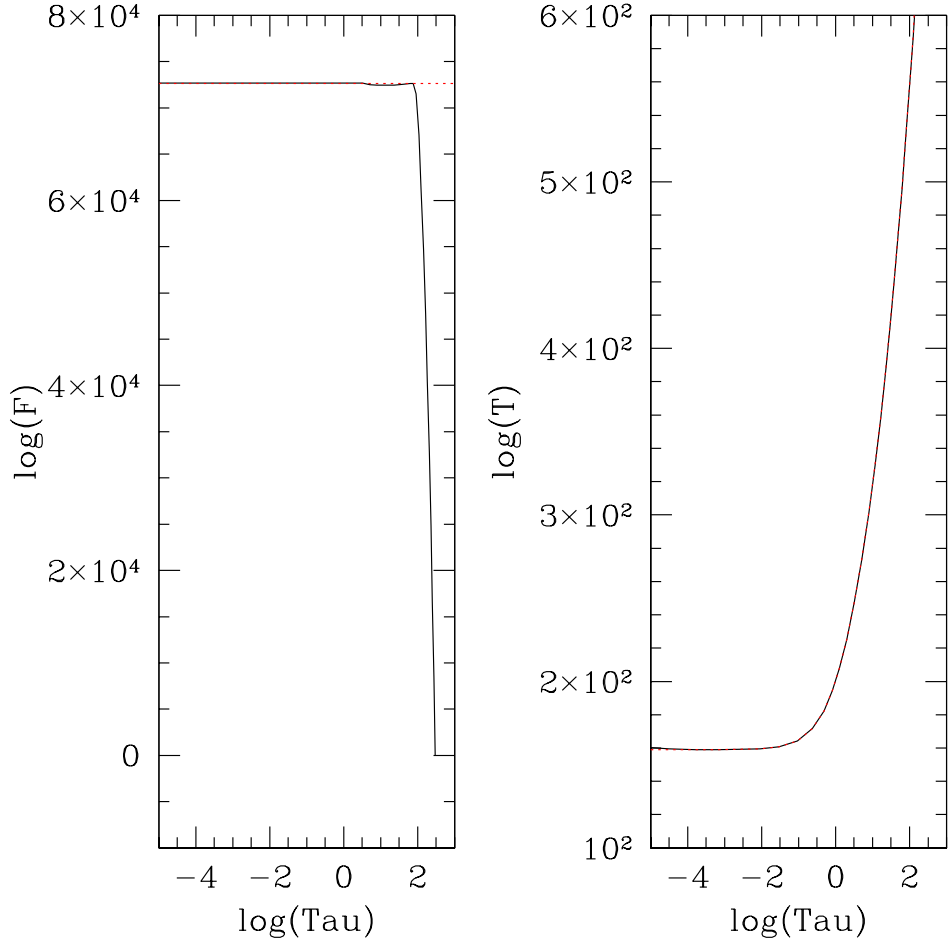} \caption{In this test, when
$\tau<100$, the viscous heating (2D) is turned off so that the flux
is a constant there and can be compared with the gray atmosphere
structure. The left panel shows the flux through the atmosphere at
different heights, while the dotted line is $F=\sigma T_{eff}^{4}$.
The right panel shows the
 disk temperature structure at one radius from
the 2D simulation (solid line) and the theoretical gray atmosphere
temperature structure (T$^{4}$=3/4 T$_{eff}^{4}$($\tau$+2/3) (dotted
line). }\label{fig:1}
\end{figure}

\clearpage
\begin{table}
\begin{center}
\caption{2D models \label{tab1}}
\begin{tabular}{cccc}

\tableline\tableline
Labels  & inner radius (AU)& half or full disk & resolution\tablenotemark{a} \\
\tableline
A &  0.3 & half & 80 $\times$ 112\\
B &  0.3 & half & 160 $\times$ 112\\
C &  0.3 & half & 320 $\times$ 112\\
D &  0.1 & half & 80 $\times$ 128\\
E &  0.1 & half & 160 $\times$ 112\\
F &  0.1 & half & 324 $\times$ 112\\
G &  0.1 & half & 576 $\times$ 112\\
H &  0.06& half & 304 $\times$ 112\\
I &  0.06& half & 512 $\times$ 112\\
J &  0.06& full & 128 $\times$ 240\\
K &  0.06& full & 320 $\times$ 224\\
 \tableline
\end{tabular}
\tablenotetext{a}{radial resolution $\times$ vertical resolution}
\end{center}
\end{table}
\clearpage

\end{document}